# First Steps to Bistatic Focusing

Koba Natroshvili, Otmar Loffeld, Holger Nies, Amaya Medrano Ortiz
Center for Sensorsystems (ZESS), University of Siegen
Paul-Bonatz-Str. 9-11, D-57068 Siegen, Germany
Tel. + 49 271 740 2432, FAX + 49 271 740 2336
natroshvili@ipp.zess.uni-siegen.de

*Abstract*—Although this work is a bit theoretical and contains lots of derivations, it leads to very explicit practical results in bistatic focusing. Our approach is based on Loffeld's [1] bistatic formula (LBF) describing the point target's (PT) reference spectrum for arbitrary bistatic configuration. Based on various simulations the validity of LBF for both airborne and spaceborne configurations is demonstrated. Focusing for special bistatic configurations like: 'Tandem' and 'Translationally Invariant' (TI) constellations is considered (definitions after [6]). The focusing for the Tandem configuration is solved analytically. Focusing in the TI case is realized by blockwise processing. All focusing algorithms are developed in IDL and adequate simulation results are presented. In the end of the paper outlines the conceptual solution of the most difficult bistatic 'General Case' (GC) and presents some first focusing results.

*Keywords-Bistatic SAR, focusing, method of stationary phase, ISFFT, target point reference spectrum*

## I. INTRODUCTION

Bi- and multistatic SAR systems have been attracting considerably increasing interest in this decade, showing benefits like: flexibility, reduced vulnerability for military applications, ability to use multi level interferometry, possibility to reduce PRF [12], etc. All the benefits are, however, paid with the price of increased processing complexity.

In [5] the idea of bistatic focusing is considered for the Tandem configuration (transmitter and receiver are following each other on the same track with some fixed offset, with equal velocities). Initial SAR raw data is convolved with *Rocca's smile* operator. In the Tandem case this operator is only *slowly range* variant. After this convolution any monostatic processor can be employed to yield the final result. Another approach for solving the bistatic problem is offered in [6], [7]. Modified monostatic processors like: Range-Doppler, Backprojection, Omega-k processors have been suggested. These algorithms to our understanding seem to be employable only for the Tandem and TI case.

Some other publications on the bistatic focusing problem, are based on intuition, empirical insights or complex numerical methods, but the solution for the general case (GC), when transmitter and receiver move with different velocities on arbitrary trajectories is not available yet. Most of the results presented in the work are based on simulated raw data. Here we especially consider Point Targets (PTs) and groups of point targets.

## II. MODELING THE BISTATIC PROBLEM

The general case bistatic geometry is presented in Fig 1, c.f. [1]. The point target coordinates are expressed in receiver coordinates $(R_{0R}, \tau_{0R})$. $R_{0R}$ is the slant range from the PT to the receiver at the point when the receiver is at the point of closest approach (PCA) to PT. $\tau_{0R}$ is the azimuth time instant, when the receiver is at closest distance to the point target. A more detailed description is presented in [1].

The bistatic phase history is the sum of the individual phase histories of transmitter and receiver. In [1] based on the geometrical model and using vectorial representations, the point target response is first modeled in the time domain and then transformed into the range/azimuth frequency domain, yielding LBF.

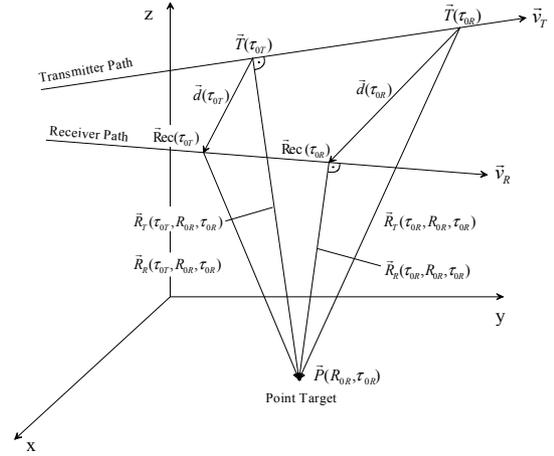

Fig 1: Bistatic Geometry

## III. PT SPECTRUM - LBF

LBF (1) is used as an initial point of future processing.

$$G_i(f, f_\tau, R_{0R}, \tau_{0R}) \cong S_i(f) \cdot w(\tilde{\tau} - \tau_{cb}) \cdot$$

$$\exp\left[-j\underbrace{(\phi_T(\tilde{\tau}_T, f_\tau) + \phi_R(\tilde{\tau}_R, f_\tau))}_{\Psi_1(f, f_\tau)}\right] \cdot \text{(Quasi monostatic term)}$$

$$\exp\left[-\underbrace{\frac{j}{2} \cdot \frac{\ddot{\phi}_T(\tilde{\tau}_T) \cdot \ddot{\phi}_R(\tilde{\tau}_R)}{\ddot{\phi}_T(\tilde{\tau}_T) + \ddot{\phi}_R(\tilde{\tau}_R)} \cdot (\tilde{\tau}_T - \tilde{\tau}_R)^2}_{\Psi_2(f, f_\tau)}\right] \quad (1)$$

$$\text{Amplitude factor} \rightarrow \cdot \frac{\sqrt{2\pi}}{\sqrt{\ddot{\phi}_T(\tilde{\tau}_T) + \ddot{\phi}_R(\tilde{\tau}_R)}} \cdot \exp\left(-j\frac{\pi}{4}\right)$$

$$a_0 = \frac{\left(\vec{R}_R(\tau_{0R}, R_{0R}, \tau_{0R}) + \vec{d}(\tau_{0R})\right) \cdot \vec{v}_T}{v_T^2} = \tau_{0T} - \tau_{0R} \quad (2)$$

$$a_2 = \sqrt{\left(\vec{e}_R(\tau_{0R}, R_{0R}, \tau_{0R}) + \frac{\vec{d}(\tau_{0R})}{R_{0R}}\right)^2 - v_T^2 \cdot \frac{a_0^2}{R_{0R}^2}} = \frac{R_{0T}}{R_{0R}}$$

$\Psi_1(f, f_\tau)$ is a quasi monostatic term. $\Psi_2(f, f_\tau)$ is a bistatic deformation phasor. $S_l(f)$ is the transmitted chirp spectrum in the low pass domain. $w(\tilde{\tau} - \tau_{cb})$ is an azimuth window expressing the time, when PT is in the beamwidth of transmitter and receiver simultaneously. It is equivalent to a rectangular window around the bistatic Doppler centroid frequency with the window width expressing the azimuth or bistatic Doppler bandwidth. Azimuth center frequency (bistatic Doppler centroid) and bandwidth estimates are presented in [2]. Estimation methods for the Doppler centroid frequency will be given in [8]. Very important parameters describing the 'bistatic grade' are $a_0$ and $a_2$ (2). $a_0$ expresses the azimuth time difference between the azimuth times of the PCAs of transmitter and receiver. $a_2$ gives the ratio of slant ranges at the PCAs.

## IV. VALIDITY OF THE APPROACH

LBF was derived using the Method of Stationary Phase (MSF). This method could be easily applied to the individual monostatic phase histories of either transmitter or receiver. In the bistatic case, where the phase history is sum of the individual phase histories of transmitter and receiver, the use of MSF should be generally doubtful. But as the sum of two quadratic terms is still a quadratic term, the individual phase histories were first expanded in two individual second order Taylor series expansions around the individual points of stationary phase and then combined to a common second order Taylor series around the common bistatic stationary point. This particular step of derivation caused some discussions and initial disagreement in the SAR community. In reaction some constraints regarding LBF's validity have been given in [1]. But clearly the best method to demonstrate the validity of the approach is by showing focusing results.

3D bistatic simulations were performed in IDL. These simulations give the possibility to generate bistatic SAR raw data for both space- and airborne cases with arbitrary parameters and configurations.

As a first test only one point target was simulated and the raw data focused with LBF. Various configurations were considered, all showing extremely well focused results. Just for comparison the same kind of focusing was done with a typical monostatic formula. It gave strongly degraded results.

Table 1 shows the parameters of one specific airborne GC simulation (even though the magnitudes of velocities of transmitter and receiver are equal the trajectories are not parallel).

TABLE 1: AIRBORNE GENERAL CASE

| Simulation parameters | Transmitter | Receiver |
|---|---|---|
| Speed of airplanes | 98m/s | 98m/s |
| Pulse duration | 3us | |
| CF | 10.17GHz | |
| Bandwidth | 20MHz | |
| PRF | 1250Hz | |
| Squint angle | 0° | 0° |
| Off Nadir angle | 42° | 52° |
| Max distance between airplanes | 1029m | |
| Min distance between airplanes | 1000m | |
| Distance of closest approach | 3893 | 4603 |
| a0 | -1.37s | |
| a2 | 1.18 | |

Fig 2 shows the focusing results: While bistatic focusing is correct (left column), the right column shows clearly visible defocusing effect in the case of applying the monostatic formula. The PT is blurred both in azimuth and range direction.

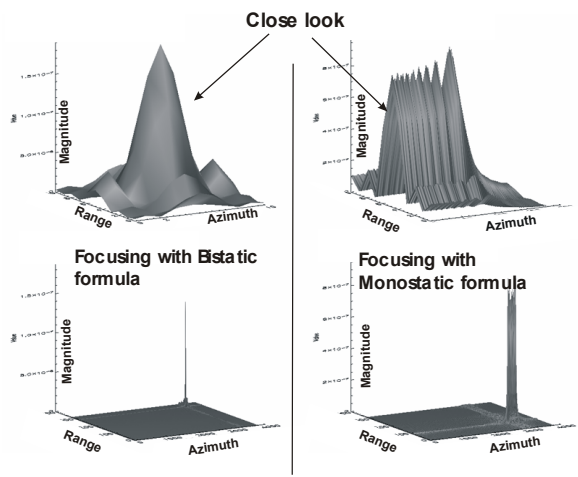

Fig 2: Airborne case focusing with Bistatic (left column) and Monostatic Formulas (right column), upper row is a closer look

## V. EXTENTION TO FOCUSING OF COMPLETE SCENES

As the raw data spectrum is sum of reflected signals from all PTs, the raw data spectrum could be expressed as an integral:

$$\iint G_l(f, f_\tau, R_{0R}, \tau_{0R}) dR_{0R} d\tau_{0R} \quad (2)$$

The aim of focusing is to somehow invert the integral (2) and deduce $\sigma(R_{0R}, \tau_{0R})$ - the bistatic backscattering coefficient.

In formula (1), both quasi monostatic and bistatic deformation phasors are as range as well as azimuth time variant. This makes the GC focusing a not trivial problem. Besides, as $(R_{0R}, \tau_{0R})$ are coordinates, all time variant parameters in (1) have to be expressed in given coordinates. First of all this refers to $R_{0T}, \tau_{0T}$.

There are, however, some very specific configurations, where LBF formula can be simplified. The first very important case is the Tandem configuration.

### A. Tandem Configuration

For the Tandem case LBF simplifies to:

$$G_l(f, f_\tau, R_{0R}, \tau_{0R}) \cong \sigma(R_{0R}, \tau_{0R}) \cdot S_l(f) \cdot \underbrace{\frac{v\sqrt{(f+f_0)}}{\sqrt{cR_{0R}}}}_{\text{Amplitude factor}} \cdot$$

$$\exp\left(-j\frac{\pi}{4}\right) \cdot w(\tilde{\tau} - \tau_{cb}) \cdot$$

$$\underbrace{\exp\left[-j\pi\left((2\tau_{0R} + a_0)f_\tau + \frac{4\pi}{c}R_{0R}F^{1/2}\right)\right]}_{\Psi_1(f, f_\tau)} \quad (3)$$

$$\cdot \underbrace{\exp\left[-j\frac{\pi d^2 F^{3/2}}{c(f+f_0)^2 2R_{0R}}\right]}_{\Psi_2(f, f_\tau)}$$

The quasi monostatic term is converted to an exactly monostatic term. $d$ is the baseline between the vehicles.

Both, bistatic and monostatic terms only vary over slant range and are invariant with respect to azimuth time. Simulations show that the *bistatic phase term is comparably negligibly varying* in comparison with the monostatic phase term. The same observation was made in [5]. Based on that observation we can linearize the bistatic term in the following way:

$$\frac{1}{R_{0R}} \approx \frac{1}{R_{0R}(\min)} - \frac{r}{R_{0R}^2(\min)} \quad \text{where: } R_{0R} = R_{0R}(\min) + r \quad (4)$$

Besides, because of the weak dependency of $F^{\frac{1}{2}}$ on $f_\tau$ the following substitution (5) is possible:

$$F^{\frac{3}{2}} \approx (f + f_0)^2 \cdot F^{\frac{1}{2}} \quad (5)$$

After the preceding modifications we combine the bistatic and monostatic phase terms. By this the bistatic focusing task is transformed to a modified monostatic processing task. As monostatic processor the Inverse Scaled FFT (ISFFT) [3] is used. Simulations prove our derivations. The simulation parameters are given in TABLE 2 and the corresponding focusing results in Fig 3. The scene consists of 2 PTs, separated by a distance of 10km in range direction.

TABLE 2, SPACEBORNE TANDEM CASE

| Simulation parameters | Transmitter | Receiver |
|---|---|---|
| Speed of satellites | 7000m/s | 7000m/s |
| Pulse duration | 8.5us | |
| CF | 5.16GHz | |
| Bandwidth | 40MHz | |
| PRF | 2500Hz | |
| Squint angle | 0.1° | 0.1° |
| Off Nadir angle | 45° | 45° |
| Distance between satellites (constant) | 1000m | |
| Distance of closest approach | 3893 | 4603 |

The magnitudes of the 1st PT and 2nd PT were deliberately chosen different to distinguish the directions. The distance between the focused peaks is in accordance with the location of the TPs.

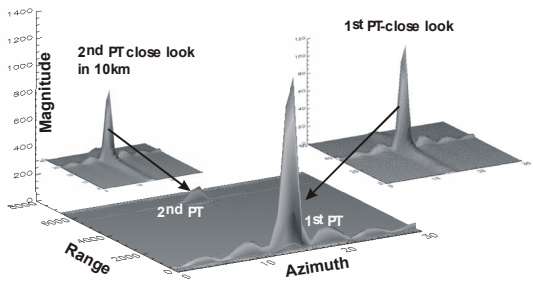

Fig 3: Focusing result for Tandem configuration

### B. Tranlationary Invariant Configuration

Another very important bistatic constellation is the Translationally Invariant (TI) configuration [6], where transmitter and receiver move on parallel tracks with equal velocities. Still, LBF only depends on slant range and is invariant with respect to azimuth time. After some rearrangements formula (1) simplifies to:

$$G_l(f, f_\tau, R_{0R}, \tau_{0R}) \cong \sigma(R_{0R}, \tau_{0R}) \cdot S_l(f) \cdot \frac{v\sqrt{(f+f_0)}}{\sqrt{c \cdot (R_{0R}+R_{0T})_{aver}}} \cdot$$

$$\exp\left(-j\frac{\pi}{4}\right) \cdot w(\tilde{\tau} - \tau_{cb}) \cdot \quad \text{Amplitude factor}$$

$$\exp\left[-j\pi\underbrace{\left((2\tau_{0R}+a_0)f_\tau + \frac{4\pi}{c}\frac{R_{0R}+R_{0T}}{2}F^{1/2}\right)}_{\Psi_1(f,f_\tau)}\right] \quad \text{Quasi monostatic term} \quad (6)$$

$$\cdot \exp\left[-j\frac{2\pi}{c}\underbrace{\frac{v^2}{(f+f_0)^2}\frac{F^{\frac{3}{2}}}{R_{0R}+R_{0T}}\left(a_0 - f_\tau\frac{c}{2v^2}\frac{R_{0T}-R_{0R}}{F^{\frac{1}{2}}}\right)}_{\Psi_2(f,f_\tau)}\right]^2 \quad \text{Bistatic term}$$

From (6) it is clear that basically 2 problems have to be solved:

- Compensation of bistatic term;
- Expression $R_{0T}$ over $R_{0R}$;

Due to the inequality of transmitter and receiver slant ranges (as opposite to the Tandem case) processing for the TI case is slightly different from the Tandem case. While in Tandem case the bistatic acquisition could be understood as a monostatic one located in middle of the baseline, the situation is different for the TI case. This is easily seen from (6), where backscattering coefficient coordinates are expressed over $R_{0R}$ and not over $(R_{0R}+R_{0T})/2$. We propose range blockwise processing- leading to a modified TI ISFFT algorithm:

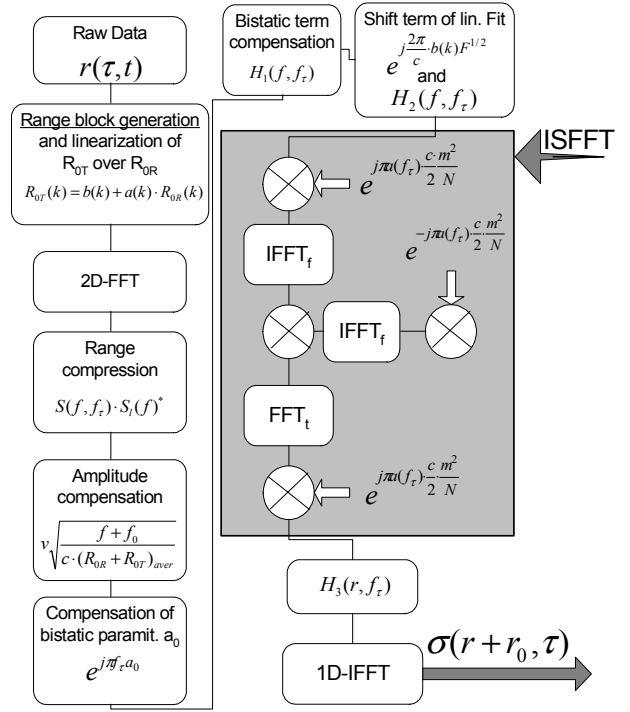

Fig 4: Modified TI ISFFT algorithm

- Range blocks are generated;
- In each block the bistatic term is averaged and compensated;
- In the quasi-monostatic term, $R_{0T}$ is linearized over $R_{0R}$;
- ISFFT processing is used;

The schematic diagram of whole algorithm is shown in Fig 4.

*1) Results with simulated raw data*

Now the scene consists of 5 equidistantly located (1km separation) PTs in range direction. The magnitudes of the PTs are weighted (brightness varies linearly over slant range). Fig 5 shows the focusing result for the TI case with similar spaceborne parameters like in TABLE 2. Each point is correctly focused and correctly located. We still have the freedom of making smaller range bins (much less then range block) and compensate bistatic terms binwise. This procedure could be quite useful in an airborne case.

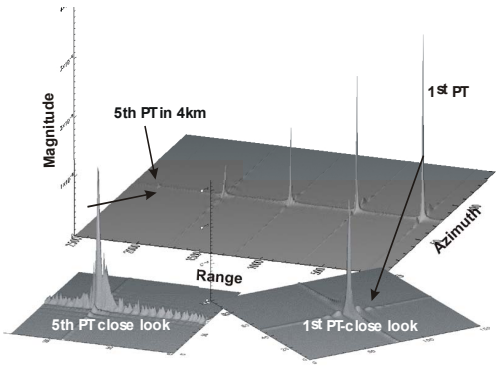

Fig 5: Focusing for Translationally Invariant configuration

*2) Results with real bistatic SAR data*

The details of focusing real TI bistatic data and a detailed description of the experiment and the results obtained will be presented in [9], [10]. The data has been provided by FGAN (German Research Establishment for Applied Natural Sciences) as a part of collaboration on bistatic SAR, which is gratefully appreciated. Here we only show a cutout of a focused result (Fig 6), to demonstrate the validity of the approach. The scene shows Oberndorf (a. Lech), the data has been acquired by FGAN's PAIMR/AER-2 systems.

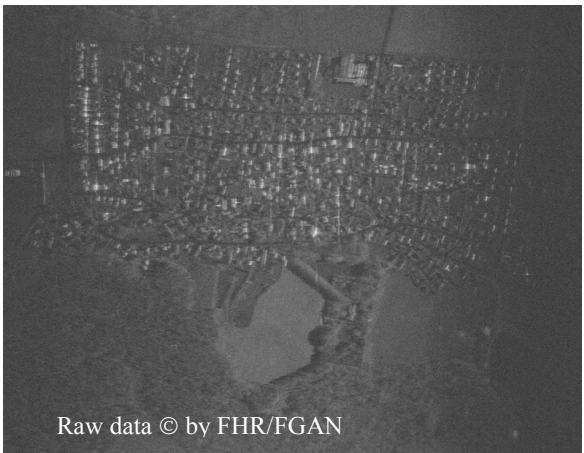

Raw data © by FHR/FGAN

Fig 6: Oberndorf (a. Lech) Bistatic SAR image

The experiment is not exactly a TI case, since the flight tracks were not strictly parallel for the whole data take. Thus parameter tuning for the processing is crucial and will be described in a joint paper [9].

## VI. FOCUSING FOR GENRAL CASE

In this case processing becomes additionally truly azimuth time dependent. The bistatic deformation phasor is still compensated blockwise with respect to range and azimuth. The individual steps are:

- Whole scene is divided in slant range - azimuth blocks;
- In these blocks $R_{0T}$ and $\tau_{0T}$ are expressed over $R_{0R}$ and $\tau_{0R}$ with linear regressions;
- Regression results are substituted in (1);
- The bistatic term is averaged and compensated in each range-azimuth block;
- Resubstituting the results into (2) the integral leads to backscattering coefficient spectrum, which is scaled and shifted in both- range and azimuth frequency directions:

$$\sigma(a_{ra}f + b_{ra}f_0, a_{az}f_\tau + b_{az})$$
$$\Downarrow \quad \Downarrow \quad \Downarrow \quad \Downarrow$$
$$f_\tau \quad f_\tau \quad const \quad f$$

This kind of result should be expected. The final aim is to express the backscattering coefficient $\sigma(r,\tau)$ compensating scaling and shift factors for both directions.

It is interesting to note that:

- Scaling factor and shift of the range frequency depend on the azimuth frequency $f_\tau$;
- The scaling factor of the azimuth frequency in the *first order approximation is constant*, but the shift depends on the range frequency $f$;

These observations allow us to separate the scaling compensation steps. First the scaling and shifts with respect to range are compensated. Later the constant azimuth scaling and variant shifts are corrected. For the moment ISFFT is used to eliminate the scaling in both directions. In the future normal Chirp Scaling could be used. In [3] the duality of these two methods is shown.

To demonstrate the validity of this approach we conducted a further simulation with spaceborne parameters, given in TABLE 3.

TABLE 3, SPACEBORNE GENERAL CASE

| Simulation parameters | Transmitter | Receiver |
|---|---|---|
| Speed of satellites | 7000m/s | 7100m/s |
| Pulse duration | 8.5us | |
| CF | 5.16GHz | |
| Bandwidth | 20MHz | |
| PRF | 1800Hz | |
| Squint angle | 0° | 0° |
| Off Nadir angle | 45° | 45° |
| Max distance between satellites | 4874m | |
| Min distance between satellites | 5100m | |

The satellites have different velocity vectors (the abs values are different and they do not move on parallel tracks), reflecting the very general case (GC). The scene consisted of 15 PTs, located on the vertexes of a matrix 3 (range column) x 5 (azimuth row), with 1 km separation in each direction. The GC bistatic algorithm was used for processing. *After compensating range scaling and shifts, the PTs look correctly focused (*Fig 7*), but a closer look shows their wrong displacements*. The effect can be explained by the fact that azimuth scaling and shifts are still to be compensated.

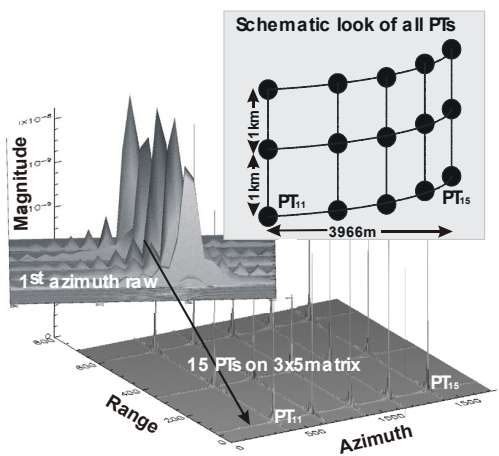

Fig 7, Focusing of 15 PTs after compensating range scaling

When regarding Fig 7 it should be kept in mind that still the <u>azimuth scaling factors</u> have not been compensated. The figure also shows a closer look at the 1st azimuth row and the schematic locations of all PTs. In the azimuth distance of 5km PTs move in range over 35m (range resolution is 7m). So we experience wrong range positioning due to uncompensated azimuth shifts.

Furthermore as a drawback of the still uncompensated azimuth scaling the PTs are incorrectly located with respect to azimuth direction: The distance between the $PT_{11}$ (1st row and 1st column) and $PT_{15}$ is 3966m instead of 4000m (which were to be expected).

As the next step, azimuth scaling and shifts are compensated. Now all PTs move to their correct positions. Fig 8 shows the final result of 2D range-azimuth bistatic focusing together with 1st azimuth row and 1st range column.

Inspecting the first range column and the first azimuth row of Fig 8 we see that now all point targets are arranged in a perfect straight line, that there is no more range or azimuth walk present. Furthermore all the point targets are quite nicely focused, finally demonstrating the validity of the approach.

### DISCUSSION AND SUMMARY

Based on simulated bistatic SAR data the validity of LBF (1) has been proven. For the special bistatic case described as the Tandem configuration, a direct and simple solution is readily obtained. Focusing of TI (translationally invariant) configuration data is done blockwise. Here the approach was verified with simulated and real data. GC processing is realised by compensating both slant range and azimuth scaling factors. All algorithms are implemented in IDL and they give very correct focusing results

### ACKNOWLEDGEMENTS

We are very grateful to F. Rocca [5] for supplying us with some interesting materials. The results presented in the paper would not have been possible without the fruitful and stimulating cooperation with FGAN's Institute for High Frequency Physics and Radar Technique (FHR), http://www.fhr.fgan.de/fhr/fhr_home_e.html, namely J. Ender, see [6], A. Brenner and I. Walterscheid [9], [10]. The authors are very grateful for this. Special thanks to our college V. Peters developing the initial bistatic simulator. We would like to gratefully acknowledge the support of DAAD's IPP program and the funding of the Ministry of Science of Northrine Westphalia.

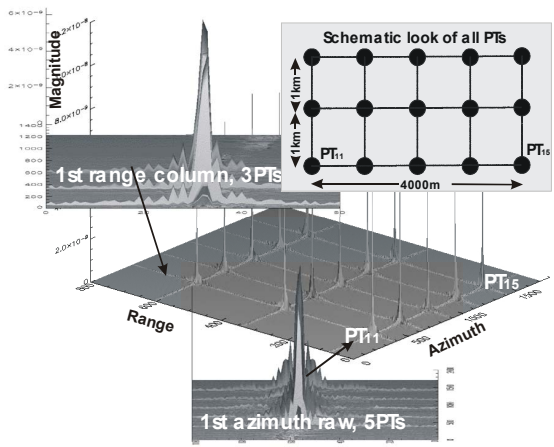

*Fig 8: Final Result of Bistatic Focusing of GC*